\documentclass{article}

\usepackage{url}
\usepackage{graphicx}

\begin{document}
        
\title{Representative elementary volume via averaged scalar Minkowski functionals} 
\author{M.V. Andreeva
\thanks{A.P. Ershov Institute of Informatics Systems of SB RAS,630090 Novosibirsk, Russia,
e-mail: maria$\_$and@mail.ru}
\and
 A.V. Kalyuzhnyuk
 \thanks{Peter the Great St.Petersburg Polytechnic University, 195251 St. Petersburg, Russia,
e-mail: Kalyuzhnyuk.AV@gazprom-neft.ru}
\and
V.V. Krutko
\thanks{OOO ``Gazpromneft NTC'', 190000 St. Petersburg, Russia,
e-mail: Krutko.VV@gazpromneft-ntc.ru}
\and 
N.E. Russkikh
\thanks{A.P. Ershov Institute of Informatics Systems of SB RAS, 630090 Novosibirsk, Russia,
e-mail: russkikh.nikolay@gmail.com}
\and
I.A.~Taimanov
\thanks{Sobolev Institute of Mathematics of SB RAS, and Novosibirsk State University, 630090, Novosibirsk, Russia, e-mail: taimanov@math.nsc.ru}
}
\date{}

\maketitle              
\begin{abstract}
Representative Elementary Volume (REV) at which the material properties do not vary with change in volume is an important quantity for making measurements or simulations which represent the whole.
We discuss the geometrical method to evaluation of REV based on the quantities coming in the Steiner formula from convex geometry. For bodies in three-dimensional space this formula gives us four scalar functionals known as scalar Minkowski functionals. We demonstrate on certain samples that the values of such averaged functionals almost stabilize for cells for which the length of edges are greater than certain threshold value $R$. Therefore, from this point of view, it is reasonable to consider cubes of volume $R^3$ as representative elementary volumes for certain physical parameters of porous medium.
\end{abstract}

\section{Introduction}

There are few notions of representative elementary volumes (REV) \cite{Hill,DW,Zhang} which have in common the condition that this is the minimal elementary volume which serves a value representative of a certain property of the whole media. Therewith it is important to mention in which respects it gives such a representation.  We address the problem of elementary volumes which represent the permeability properties of porous medium. Such a problem was considered from different points of view, for instance, in \cite{DO,SAR,Bear}.

If we know the representative elementary volume then we can perform all hydrodynamic simulations on such volumes to evaluate effective characteristics of the medium. Unlike modeling on a full-scale model, this allows us to minimize the computations of such characteristics as the Darcy coefficient.

For this reason, the challenge is to find geometric characteristics related to permeability to use them for evaluating representative elementary volumes. Recently different possible applications to estimating permeability of two- and three-dimensional porous media by using the scalar Minkowski functionals were discussed in \cite{Scholz,Armstrong,Slotte}. Therefore we propose to use the averaged scalar Minkowski functionals as such geometric characteristics. In this article we show that such functionals detect certain REVs, including the REV of porosity.

A porous medium is represented by a digital core consisting of voxels that correspond to elementary cells. Each cell corresponds to the value of the radiodensity.

\begin{figure}[h!]
        \centering
        \includegraphics[width=180px]{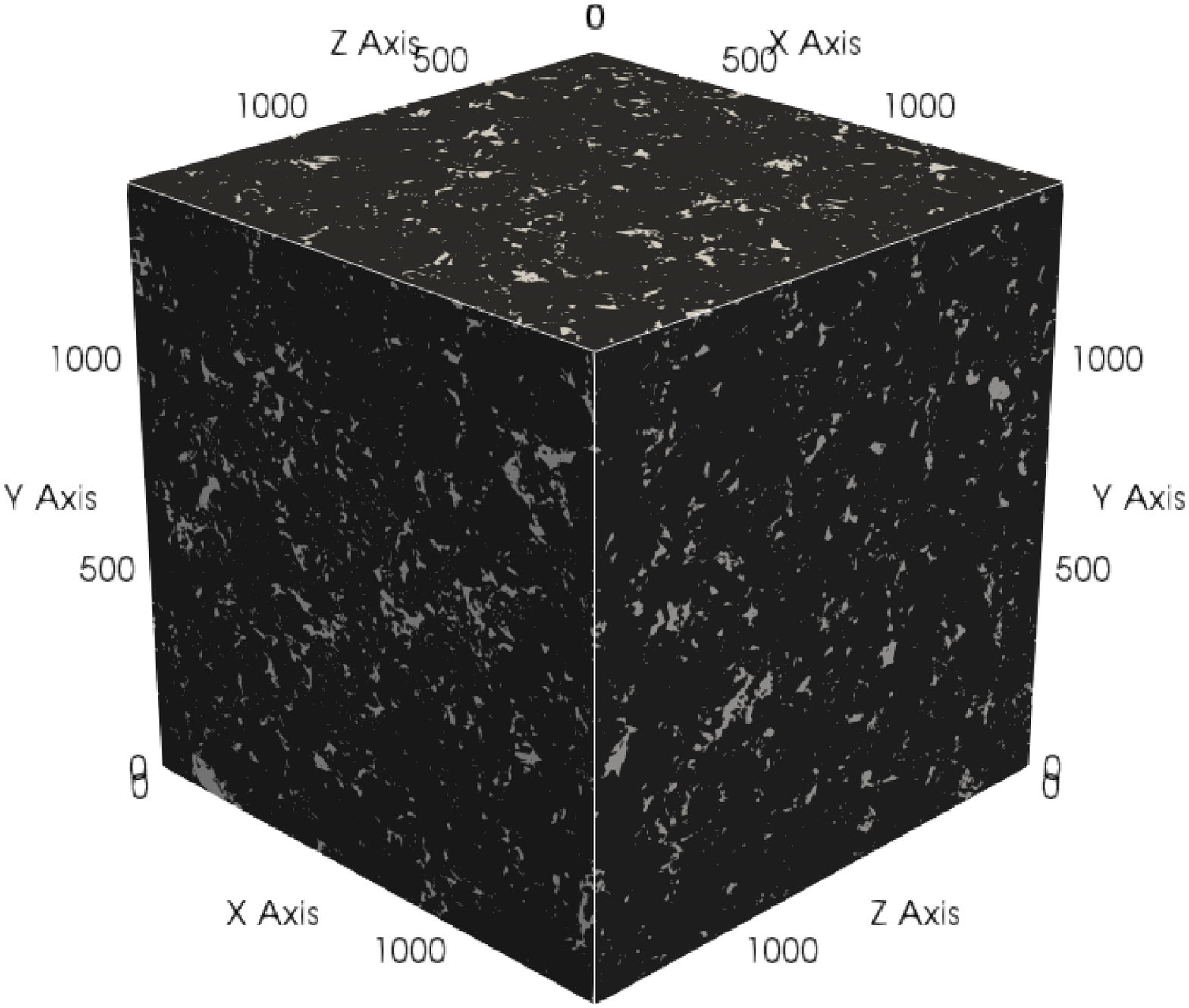}
        \center{Fig. 1. Binarized digital core. 1400x1400x1400 voxels.}
        \center{$ $}
        \label{fig:binarized_core}
\end{figure}

Picking up the excursion coefficient $\lambda, 0 < \lambda < 1$, we assume that the voxel is black if the proportion is greater than or equal to $\lambda$ and it is white otherwise. As a result we obtain a two-colored (binarized) digital core which, we assume, consists of $N \times N \times N$ black or white voxels (see Fig.~\ref{fig:binarized_core}). Each voxel corresponds to a cubic cell of size $L \times L \times L$.

We take randomly a sample voxel and correspond to it the set of nested cubes centered at it. For each such a cube $X$  we compute the averaged Minkowski functionals $W_i(Y)/\mathrm{Vol}(X)$, where $Y$ is a subset of $X$ formed by black voxels. Every such a cube consists of $k \times k \times k$ voxels and we consider the values of these values on nested cubes as functions of $k$. If there exists a constant $C$ such that for generic initial the graphs of the function, the averaged functional $W_i$, goes to the asymptotes as $k \to C$ we may consider a cube of volume $R^3$ with $R=CL$ as a candidate for a representative elementary volume for the property related to $W_i$.  

In the example considered in Section 3 we take a sample of medium with $N = 1400$ and the
elementary length $L = 1.5\, \mu m$. For sample voxels we consider the graphs of the averaged functionals on nested cubes. It appears that for all sample initial voxels and for the averaged functionals $W_0$ and $W_1$ the graphs go to the asymptote as $k \to C \approx 200$, for the averaged $W_2$ functional 
it goes to the asymptote as $k \to C \approx 100$, and for the averaged $W_3$ functional $C \approx 150$.

\section{The scalar Minkowski functionals}

Let $X$ be a convex body with a regular boundary in three-dimensional space ${\bf R}^3$ and $B$ be the ball formed by all vectors of length not greater than one. We denote by $X+\varepsilon B$ all points of the form $x+\varepsilon b$ where $x$ is a point from $X$, $b$ is a vector from $B$ and $\varepsilon$ is a positive constant:
$$
x=(x_1,x_2,x_3), \ \ b = (b_1,b_2,b_3), \ \ x+\varepsilon b = (x_1+ \varepsilon b_1, x_2+ \varepsilon b_2, x_3 + \varepsilon b_3).
$$
Since the zero vector belongs to $B$ the original set $X$ is a subset of $X+ \varepsilon B$ and moreover
$X + \varepsilon_1 B$ lies in $X +\varepsilon_2 B$ if $\varepsilon_1 \leq \varepsilon_2$.
It is also easy to check that the bodies of the form $X+\varepsilon B$ are convex.

We denote by $\mathrm{Vol}(X)$ the volume of a body $X$ in three-dimensional space. The famous Steiner formula reads that
$$
\mathrm{Vol}(X + \varepsilon B) = \sum_{k=0}^3 {{3}\choose{k}} W_k(X) \varepsilon^k.
$$
The quantities $W_k, k=0,1,2,3$, are called the $k$-th quermassintegrals, or the scalar Minkowski functionals.

The latter name demonstrates that there are their generalizations to higher ranks (for instance, vector and rank two functionals). We shall not discuss them here because the investigation of their possible applications
to our main problem is in progress.

We recall that the analog of the Steiner formula is valid for bodies in the space ${\bf R}^n$ of arbitrary dimension $n$. In such a case $\mathrm{Vol} (X + \varepsilon B)$, the $n$-th dimensional volume of $X+\varepsilon B$, is a polynomial in $\varepsilon$ of degree $n$. Since for $\varepsilon = 0$ the polynomial
gives us the volume of $X$, we have
$$
W_0(X) = \mathrm{Vol}(X).
$$
Other quantities $W_k, k >0$, give us nontrivial scalar functionals.
To guess their meanings we write down the Steiner formula for a ball $X=D_R$ of radius $R$. In this case
$X+\varepsilon B$ is a ball $D_{R+\varepsilon}$ of radius $R+\varepsilon$ and we have
$$
\mathrm{Vol}(D_R+\varepsilon B) = \frac{4\pi (R+\varepsilon)^3}{3} = \frac{4\pi R^3}{3}
+ 4\pi R^2 \varepsilon +  4 \pi R \varepsilon^2 +
\frac{4}{3}\pi \varepsilon^3.
$$

We see that these quantities have the following interpretations which, in fact,
are valid for all convex bodies with regular boundaries:
$$
W_1(X) = \frac{1}{3} \, \mathrm{Area} (\partial X),
$$
where $\mathrm{Area} (\partial X)$ is the area of the boundary $\partial X$ of $X$,
$$
W_2(X) = \frac{1}{3} \int_{\partial X} H\,dA,
$$
i.e. one third of the integral of the mean curvature $H$ (for the sphere of radius $R$, which is $\partial D_R$,
it is equal to $\frac{1}{R})$, over $\partial X$,
$$
W_3(X) = \frac{1}{3} \int_{\partial X} K\,dA,
$$
i.e. one third of the integral of the Gaussian curvature $K$, which is $K=\frac{1}{R^2}$ for $\partial D_R$,
over $\partial X$. We do not dwell here on the standard notions of the curvatures referring to textbooks on geometry
(for instance, \cite{NT}).

By continuity, the definitions of these quantities are uniquely extended to valuations of all convex bodies. This is important for us because from a digital core of a porous medium we construct a union of cubes which may intersect only by their faces or vertices. In this case every cube corresponds to a voxel from a digital core. Such unions are not necessarily convex however the notions of these functionals are uniquely extended to non-convex bodies by using the additivity property:
$$
W(X \cup Y) = W(X) + W(Y) - W(X\cap Y).
$$
Therefore, the quermassintegrals (of the scalar Minkowski functionals) are defined for all unions of cubes which are naturally constructed from digital cores of porous media.

For instance, for bodies with piece-wise linear triangulated boundaries the explicit formulas for computations
in terms of combinatorial data are given in \cite{SMKSBHM}.

The functional $P(Y) = 1-\frac{W_0(Y)}{\mathrm{Vol}(X)}$ measure the volume of the space complemented to the media $Y \subset X$ and it is equal to the porosity of $Y$.

Certain topological characteristics of the media, i.e. the Betti numbers of porous media (in particular, oil and gas reservoirs) weighted by volumes were considered in \cite{Betti}.
The functional $W_3$ is expressed in terms not of the media but of its boundary as follows:
$W_3(Y) = \frac{2\pi}{3} \chi (\partial Y)$,
where the Euler characteristic of $\partial Y$ is the alternated sum of the Betti numbers $b_i$ of $\partial Y$:
$\chi (\partial X) = b_0 - b_1 + b_2$.

\section{Evaluation of REV}

The method of evaluating REV was sketched in Introduction and it is as follows.
Take a digital core which consists of $N \times N \times N$ voxels such that every voxel corresponds to the elementary cubic cell of size $L \times L \times L$.

\begin{itemize}
\item
take randomly a sample voxel and correspond to it the set of nested cubes centered at it;

\item
for each such a cube $X$  we compute the averaged Minkowski functionals $W_i(Y)/\mathrm{Vol}(X)$, where $Y$ is a subset of $X$ formed by black voxels;

\item
every such a cube consists of $k \times k \times k$ voxels and we consider the values of these values on nested cubes as functions of $k$;

\item
given sufficiently many initial sample voxels, if the graphs of these functions go to the asymptotes as $k \to C$ we consider a cube of volume $R^3$ with $R=CL$ as a representative elementary volume.
\end{itemize}

As an example we consider a sample consisting of $1400 \times 1400 \times 1400$ voxels ($N=1400$) such that every voxel
corresponds to a cube of size $1.5 \,\mu m \times 1.5 \mu m \times 1.5 \mu m$. We present below the graphs of the averaged Minkowski functionals. For every family of nested cubes  the values of the averaged functionals are given by a certain (colored) line (see Fig. 2--5).
\begin{figure}[h!]
    \begin{minipage}[t]{0.49\textwidth}
        \includegraphics[width=\textwidth]{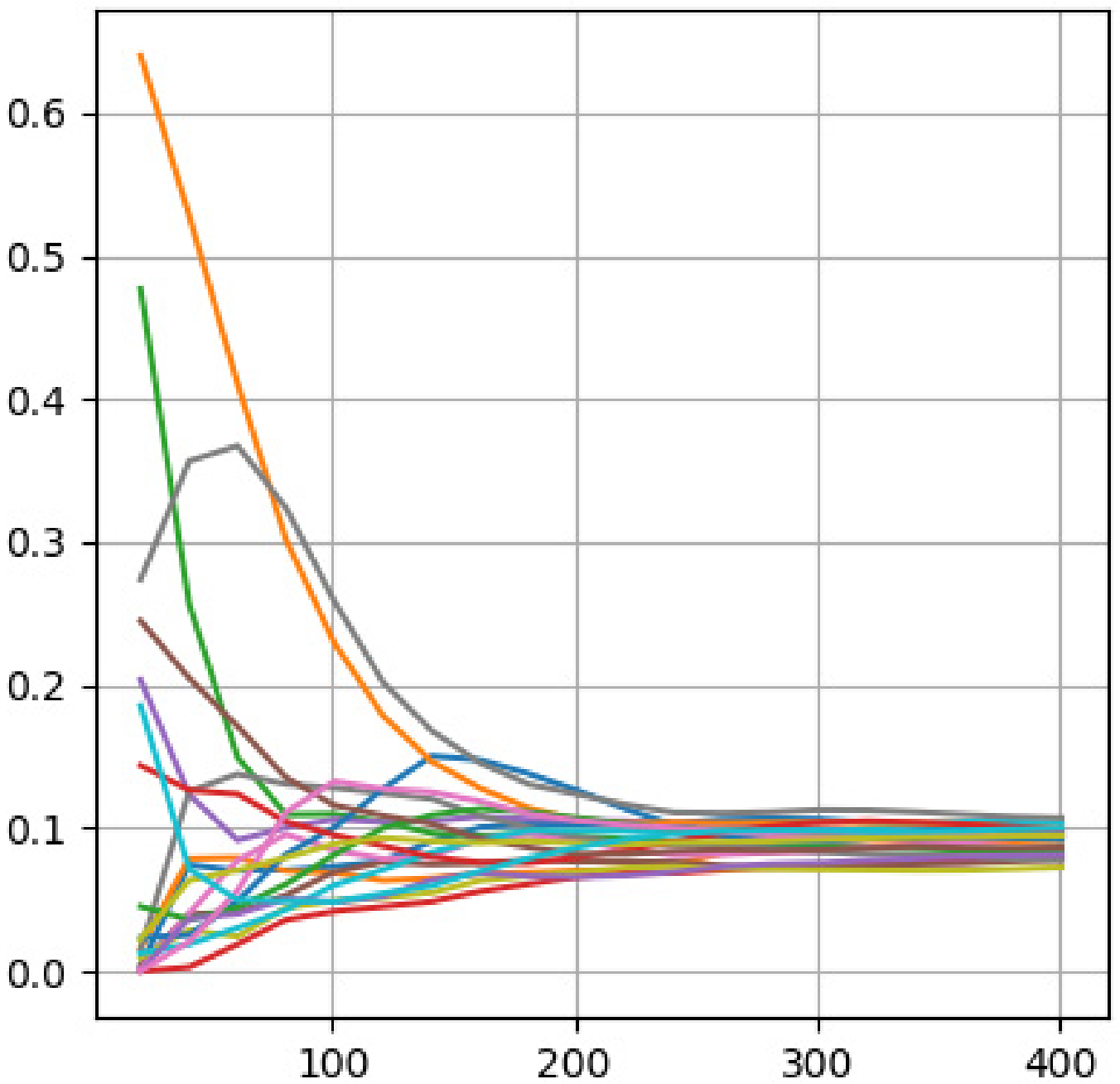}
        \center{Fig. 2. $W_0$. }
        \center{$ $}
    \end{minipage}
    \begin{minipage}[t]{0.49\textwidth}
        \includegraphics[width=\textwidth]{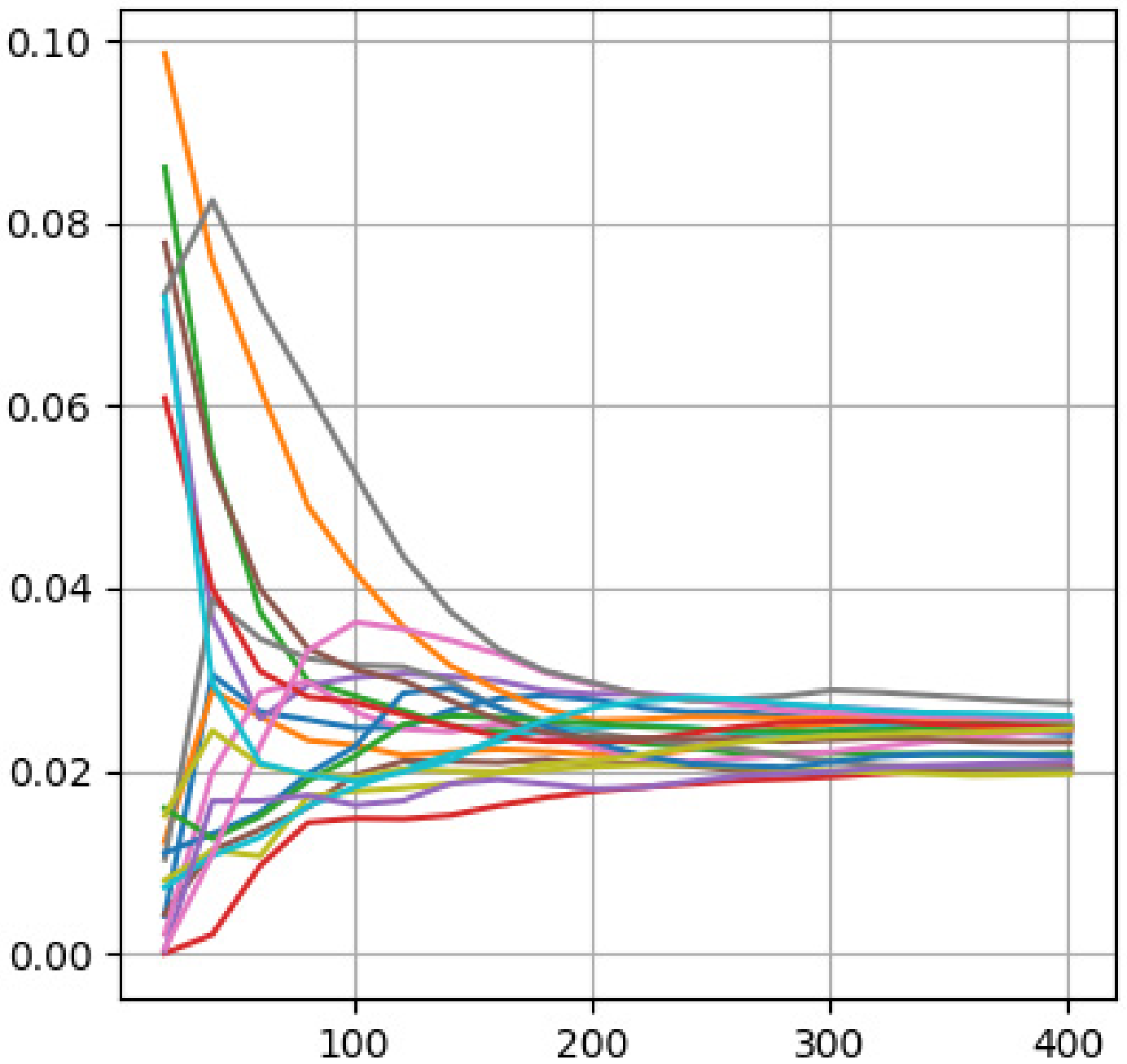}
        \center{Fig. 3. $W_1$}
    \end{minipage}

    \begin{minipage}[b]{0.49\textwidth}
        \includegraphics[width=\textwidth]{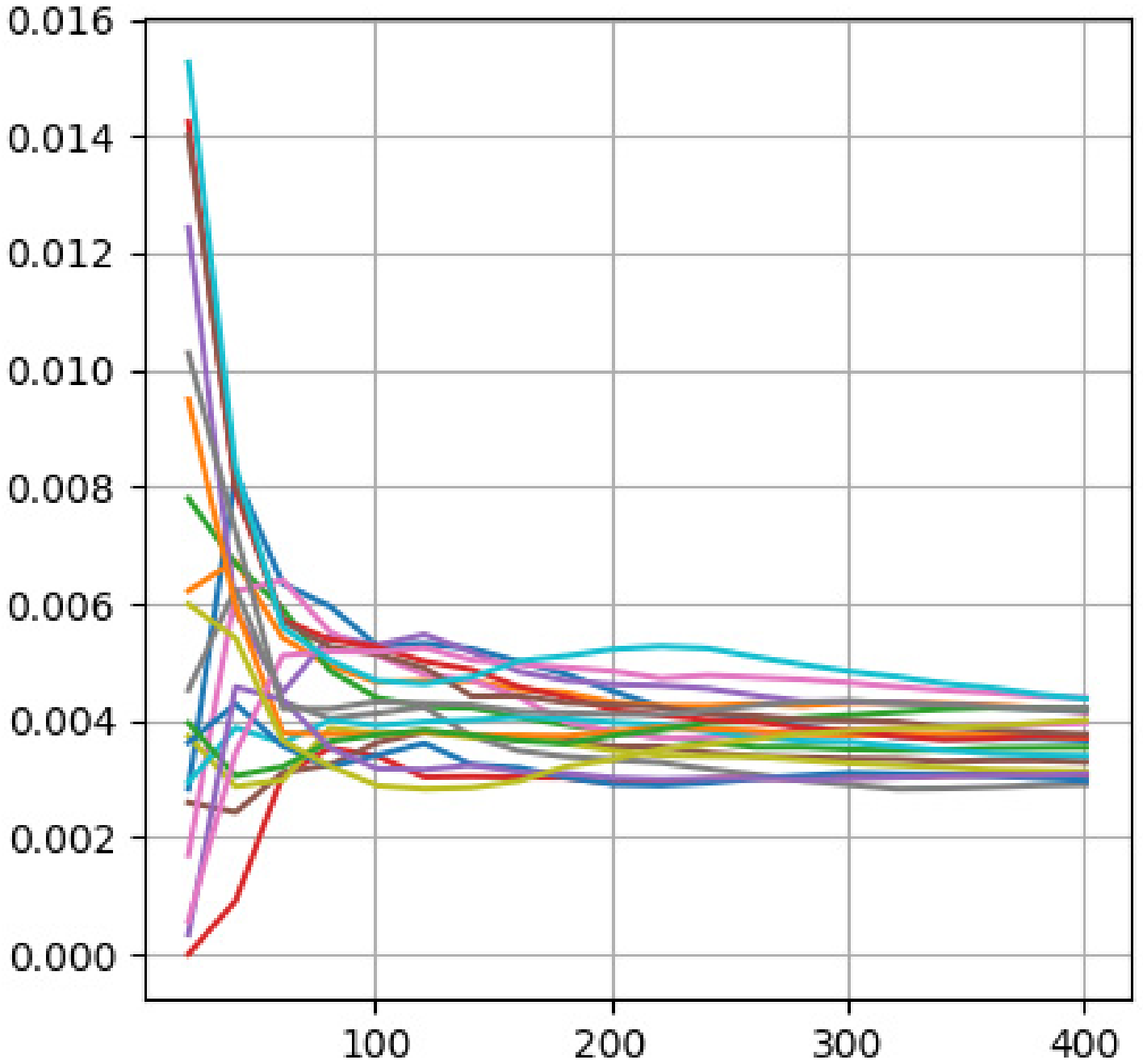}
        \center{Fig. 4. $W_2$}
    \end{minipage}
    \begin{minipage}[b]{0.49\textwidth}
        \includegraphics[width=\textwidth]{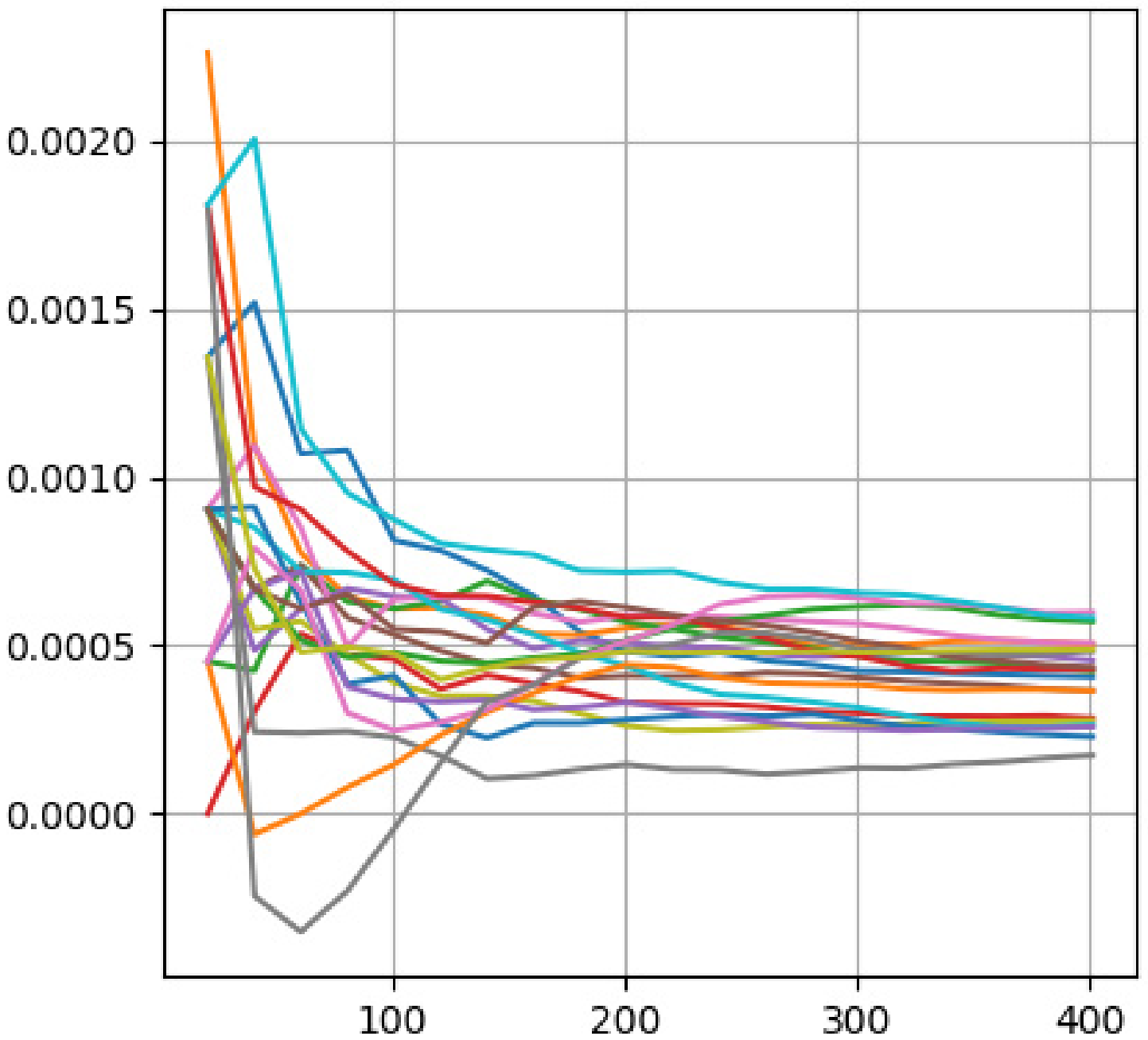}
        \center{Fig. 5. $W_3$}
    \end{minipage}
\end{figure}

We see that for the functionals which are the averages of $W_0$ and $W_1$  and for all initial voxels the graphs go to the asymptotes as $k \to C \approx 200$. For $W_2$ the graph goes to the asymptote as
$k \to C \approx 100$, and for $W_3$ we have $C \approx 150$.

{\sl This allows us to consider cubes of volume $R^3$ with $R = CL$ and $L = 1.5\,\mu m$ as the representative elementary volumes of this medium for the physical parameters related to $W_0, W_1, W_2$ and $W_3$, where $C \approx 200$ for $W_0$ and $W_1$, $C \approx 100$ for $W_2$, and 
$C \approx 150$ for $W_3$.
}

\vskip6mm

{\sc Final remarks.}

We conclude that 

{\sl the averaged Minkowski functionals have a tendency to go to asymptotes at some volumes}.

Therefore the proposed method leads to reasonable candidates for the representative elementary volumes for certain properties of the media using geometric characteristics of binarized digital core model.  Applying this procedure to $W_0$ we obtain the REV for porosity. 

We need to compare the results with the others obtained by conventional methods of REV evaluation, such as computational fluid dynamics simulations, to understand their meaning . Since the averaged functionals stabilize at different scales, the corresponding physical parameters have to be different. To our opinion it is worth to extend the study to Minkowski tensors which in addition reflect the anisotropy of media. 
These ideas may be helpful for finding REV for permeability. For instance, it was demonstrated in  \cite{Zhang,MBB} by using numerical simulation that for certain materials the spatial scales of REVs for 
permeability are approximately 1.5--2 times larger than those for porosity.

We are also left to understand how the results obtained by the proposed method vary with the excursion coefficient  $\lambda$.

\end{document}